\documentclass[a4wide,11pt]{article}
\pdfoutput=1 

\usepackage{jheppub}

\usepackage{graphicx, epstopdf}

\usepackage{amsmath, amsthm, amsfonts, amssymb, mathtools, bm, bbm}
\usepackage{empheq, physics, slashed}

\allowdisplaybreaks

\def\MSbar{\overline{\mathrm{MS}}}

\title{A note on quark and gluon energy-momentum tensors}

\author[a,b]{Taushif Ahmed,}
\affiliation[a]{Dipartimento di Fisica and Arnold-Regge Center, Universit\`a di Torino, and INFN, Sezione di Torino, Via Pietro Giuria 1, I-10125 Torino, Italy}
\affiliation[b]{Institut f\"ur Theoretische Physik, Universität Regensburg,\\ D-93040 Regensburg, Germany}
\emailAdd{taushif.ahmed@ur.de}
\author[c]{Long Chen,}
\affiliation[c]{School of Physics, Shandong University, Jinan, Shandong 250100, China}
\emailAdd{longchen@sdu.edu.cn}
\author[d]{Micha\l{} Czakon}
\affiliation[d]{Institut f\"ur Theoretische Teilchenphysik und Kosmologie, RWTH Aachen University,\\ D-52056 Aachen, Germany}
\emailAdd{mczakon@physik.rwth-aachen.de}

\abstract{We discuss the constraints on quark and gluon energy-momentum tensors in QCD that follow from the requirement of Renormalisation-Group invariance of the traces of these operators. Our study covers the most general form of the latter traces, while the energy-momentum tensors themselves are only subjected to very mild constraints. We derive Renormalisation-Group equations for the two finite independent functions of the strong coupling constant and renormalisation scale of minimal subtraction which completely define the energy-momentum tensors. We demonstrate that previously proposed definitions of the renormalized quark and gluon energy-momentum tensors are special cases of our results assuming no explicit dependence on the renormalisation scale. Finally, we present $\MSbar$-renormalised quark and gluon energy-momentum tensors at four-loop order.}

\preprint{P3H-22-088, TTK-22-28}

\begin{document}
\maketitle
\flushbottom

\section{Introduction}

The classical rank-two symmetric energy-momentum tensor (EMT) describes the density and flux of energy and momentum of a physical system. For a closed system, its four-vector components are the conserved Noether currents associated with spacetime-translation invariance up to separately conserved terms that yield the symmetric form. The density and flux of energy and momentum are the sources of the gravitational field in Einstein's field equations of General Relativity. The EMT is thus indispensable in order to describe the motion of gravitating systems.

The quantum EMT has attracted substantial attention after the discovery of the so-called trace anomaly~\cite{Crewther:1972kn,Chanowitz:1972vd,Chanowitz:1972da} in Quantum Field Theories. This anomaly is closely connected to the broken scale invariance due to quantum corrections as reflected in Callan-Symanzik equations~\cite{Callan:1970yg,Symanzik:1970rt,Coleman:1970je}. Following the early classical works on the trace anomaly in gauge field theories \cite{Freedman:1974gs,Freedman:1974ze,Adler:1976zt,Collins:1976yq,Nielsen:1977sy,Shifman:1978zn,Spiridonov:1984br} and especially in Quantum Chromodynamics (QCD), there has been interest in the connection of the anomaly with the origin and the decomposition of the nucleon mass, see e.g.\ Refs.~\cite{Ji:1994av,Ji:1995sv,Lorce:2015lna,Roberts:2016vyn,Lorce:2017xzd,Hatta:2018ina,Hatta:2018sqd,Rodini:2020pis,Metz:2020vxd,Lorce:2021xku,Ji:2021mtz,Kharzeev:2021qkd,Yang:2018nqn,Liu:2021gco,Ji:2021pys,Sun:2020ksc,He:2021bof} and references therein. Particular attention has been devoted to the proton mass, as this is unarguably one of the most fundamental and intriguingly quantities in particle physics. Interestingly, experiments proposed at the Jefferson Laboratory \cite{Dudek:2012vr} and future Electron-Ion Collider~\cite{Accardi:2012qut}, such as the near-threshold photo-production of $J/\psi$ in electron-hadron/nucleon scattering \cite{Kharzeev:1995ij,Kharzeev:1998bz,Hatta:2018ina,Kharzeev:2021qkd}, as well as the currently envisaged EicC project \cite{Anderle:2021wcy} via the $\Upsilon p$ near-threshold scattering, could shed light on the origin of the hadron/nucleon mass.

Since the first systematic decomposition of the EMT of QCD and the related decomposition of hadron masses \cite{Ji:1994av,Ji:1995sv}, commonly referred to as Ji's  decomposition scheme, a multitude of alternatives have appeared in the literature \cite{Lorce:2017xzd,Hatta:2018sqd,Rodini:2020pis,Metz:2020vxd,Lorce:2021xku,Ji:2021mtz,Yang:2018nqn,Liu:2021gco}. One reason for the different proposals is the fact that the partition of the trace of the EMT depends on the regularisation and renormalisation scheme, see e.g. Refs.~\cite{Ji:1995sv,Hatta:2018sqd,Makino:2014taa,Lorce:2021xku}. In the present publication, we will also be mainly concerned with the operator renormalisation used to define quark and gluon EMTs. Our starting point are Refs.~\cite{Hatta:2018sqd,Tanaka:2018nae}, where QCD is regularised dimensionally, while not all operators are renormalised by minimal subtraction. Our goal is to obtain well-defined finite quark and gluon EMTs that have Renormalisation-Group (RG) invariant traces. This last property is particularly useful in view of applications to the proton mass, since perturbation theory can barely be used at the relevant renormalisation scale. RG invariance removes the large scale dependence otherwise present in the expectation values of the traces of the quark and gluon EMTs. Our work extends previous studies to the most general case assuming only non-renormalisation of the total EMT. We also discuss a minimal correction of the quark and gluon EMTs defined in the $\MSbar$ scheme to render their traces RG invariant. Finally, we extend the previous three-loop $\MSbar$ results for the quark and gluon contributions to the proton and pion mass \cite{Tanaka:2018nae} to the four-loop level and compare them with their RG-invariant counterparts.

The article is organized as follows. In the next Section, we discuss the renormalisation of the physical operators occurring in the definition of the EMT. We follow the main argument of Refs.\cite{Hatta:2018sqd,Tanaka:2018nae}, but adapt the notation and simplify several steps of the derivations. Furthermore, we pay particular attention to the $\MSbar$ scheme in point~\ref{sec:MSbarConstraints}. In Section \ref{sec:RGIqgEMTtraces}, we define the individual renormalised quark and gluon EMTs and discuss the general form of their RG-invariant traces. We also present a minimal modification of $\MSbar$-scheme EMTs to obtain RG-invariant traces in point~\ref{sec:minimalRGIqgEMTtraces}. Our four-loop results for expectation values of $\MSbar$-scheme EMT traces in proton and pion states are presented in Section~\ref{sec:FourLoopResults} and compared to their RG-invariant counterparts. We conclude in Section~\ref{sec:conclusions}.

\section{Renormalisation of physical-basis operators}

This Section follows Refs.~\cite{Hatta:2018sqd,Tanaka:2018nae} to a large extent, although we modify the notation and attempt to simplify the argumentation.

The physical symmetric EMT of QCD, obtained according to the Belinfante-Rosenfeld~\cite{belinfante1939spin,belinfante1940current,rosenfeld1940energy} method, is given by:
\begin{equation}
    \label{eq:EMT}
    \Theta^{\mu\nu} \equiv -F^{a\,\mu}{}_{\rho} F^{a\,\nu\rho} + \frac{1}{4} g^{\mu\nu} F^{a}_{\rho\sigma}\,F^{a\,\rho\sigma} + \frac{i}{4} \sum_{q} \bar{q} \, \big( \gamma^{\mu} \overleftrightarrow{D}^{\nu} + \gamma^{\nu} \overleftrightarrow{D}^{\mu} \big) \, q \; ,
\end{equation}
with field-strength tensor:
\begin{equation}
    F^a_{\mu\nu} \equiv \partial_{\mu} A_{\nu}^{a} - \partial_{\nu} A_{\mu}^{a} + g_s \,  f^{abc} A_{\mu}^{b} A_{\nu}^{c} \; ,
\end{equation}
and covariant derivative:
\begin{equation}
    \overleftrightarrow{D}_\mu \equiv D_\mu -\overleftarrow{D}_\mu \; , \qquad D_\mu = \partial^{\mu} - ig_s A_{\mu}^{a} T^a \; .
\end{equation}
$A_\mu^a$ is the gluon field, $q$ is a quark field of mass $m_q$, $g_s$ is the strong coupling constant, $T^a$ are hermitian SU($N_c$), $N_c = 3$, generators, and $f^{abc}$ are real structure constants normalised according to $\comm{T^a}{T^b} =i f^{abc} T^c$. Depending on the quantisation, the actual symmetric EMT derived from the quantised Lagrangian with gauge-fixing terms contains additional terms that do not contribute in matrix elements of physical on-shell states. In linear gauges, for instance, these additional terms amount to a BRS (Becchi-Rouet-Stora) variation of a local operator \cite{Nielsen:1977sy,Collins:1976yq} (see also application at the three-loop order in Ref.~\cite{Zoller:2012qv}). In the present work, we will only be concerned with Eq.~\eqref{eq:EMT} which is necessary for physical applications.

The EMT Eq.~\eqref{eq:EMT} is conveniently expressed in a basis of local operators:
\begin{align}
    \mathcal{O}^{\mu\nu}_1 &\equiv -F^{a\,\mu}{}_{\rho} F^{a\,\nu\rho} \; , \qquad
    &\mathcal{O}^{\mu\nu}_2 &\equiv g^{\mu\nu} F^{a}_{\rho\sigma}\,F^{a\,\rho\sigma} \equiv g^{\mu\nu} \mathcal{O}_F \; , \\[.2cm]
    \mathcal{O}^{\mu\nu}_3 &\equiv \frac{i}{4} \sum_{q} \bar{q} \, \big( \gamma^{\mu} \overleftrightarrow{D}^{\nu} + \gamma^{\nu} \overleftrightarrow{D}^{\mu} \big) \, q \; , \qquad &\mathcal{O}^{\mu\nu}_4 &\equiv g^{\mu\nu} \sum_{q} m_q \bar{q} q \equiv g^{\mu\nu} \mathcal{O}_m \; ,
\end{align}
\begin{equation}
    \Theta = \mathcal{O}_1 + \frac{1}{4} \mathcal{O}_2 + \mathcal{O}_3 \; ,
\end{equation}
where we have suppressed Lorentz indices in the last equation as we will also do below. The fourth operator, $\mathcal{O}_4$, has been introduced to accommodate the renormalisation of $\mathcal{O}_2$. Indeed, there is \cite{Nielsen:1975ph,Collins:1976yq,Nielsen:1977sy,Tarrach:1981bi}:
\begin{equation} \label{eq:OFOmRenormalisation}
    \big[ \mathcal{O}_F \big]_\mathrm{R} = Z_{FF}\big[ \mathcal{O}_F \big]_\mathrm{B} + Z_{Fm} \big[ \mathcal{O}_m \big]_\mathrm{B} \; , \qquad \big[ \mathcal{O}_m \big]_\mathrm{R} = \big[ \mathcal{O}_m \big]_\mathrm{B} \; .
\end{equation}
where $\big[ \cdot ]_\mathrm{R}$ denotes a renormalised operator with finite matrix elements, while $\big[ \cdot ]_\mathrm{B}$ is defined by replacing fields, masses and coupling constants by their bare counterparts. In Eq.~\eqref{eq:OFOmRenormalisation} and in the remainder of this publication, we neglect operators vanishing by the equations-of-motion and BRS-exact operators assuming Faddeev-Popov quantisation. The renormalisation constants $Z_{FF}$ and $Z_{Fm}$ are related in the $\MSbar$ scheme with spacetime dimension $D \equiv 4 - 2 \epsilon$ to the fundamental anomalous dimensions of the theory \cite{Spiridonov:1984br}:
\begin{equation} \label{eq:Spiridionov}
    Z_{FF} = \Big( 1 - \frac{\beta}{\epsilon} \Big)^{-1} \; , \qquad
    Z_{Fm} = -\frac{4\gamma_m}{\epsilon} Z_{FF} \; ,
\end{equation}
where $\beta$ and $\gamma_m$ are defined through:
\begin{equation}
    \dv{\ln a_s}{\ln \mu^2} \equiv -\epsilon + \beta \; , \qquad a_s \equiv \frac{\alpha_s}{4 \pi} = \frac{g_s^2}{16 \pi^2} \; , \qquad \dv{\ln m}{\ln \mu^2} \equiv \gamma_m \; .
\end{equation}

The four operators $\mathcal{O}_{1,\dots,4}$ mix under renormalisation:
\begin{equation} \label{eq:OiRenormalisation}
    \big[ \mathcal{O}_i ]_\mathrm{R} \equiv Z_{ij} \big[ \mathcal{O}_j \big]_\mathrm{B} \; .
\end{equation}
Our main {\bf assumption} that will influence the quark and gluon EMTs defined in the next Section can be stated as follows:

\vspace{.5cm}
{\centering
\fbox{
\begin{minipage}{0.965\textwidth} The renormalisation constants $Z_{ij}$ are defined in the $\MSbar$ scheme for all operators but the mixing of $\mathcal{O}_{2,4}$ into $\mathcal{O}_{1,3}$. The Laurent expansion of $Z_{ij}$, $i=1,3$, $j=2,4$ may contain terms proportional to $\epsilon^k$, $k \geq 0$.
\end{minipage}
}}

\vspace{.5cm}
\noindent
We will discuss this assumption further in Section~\ref{sec:conclusions}. The renormalisation-constant matrix has the structure:
\begin{equation} \label{eq:ZijStructure}
Z = \mqty(Z_{11} & Z_{12} & Z_{13} & Z_{14} \\[.2cm]
          0      & Z_{FF} & 0      & Z_{Fm} \\[.2cm]
          Z_{31} & Z_{32} & Z_{33} & Z_{34} \\[.2cm]
          0      & 0      & 0      & 1 ) \; ,
\end{equation}
where we have taken into account:
\begin{enumerate}
    \item $Z_{ij} = 0$ for $i = 2,4$ and $j = 1,3$ because $\mathcal{O}_{2,4}$ are proportional to scalar operators, while $\mathcal{O}_{1,3}$ are (reducible) rank-two tensor operators;
    \item $Z_{4j} = 0$ for $j = 1,2,3$ because the canonical dimension of $\mathcal{O}_{4}$ is 3, less than those of $\mathcal{O}_{1,2,3}$ ;
    \item $Z_{22} = Z_{FF}$, $Z_{24} = Z_{Fm}$, $Z_{44} = 1$ because $\mathcal{O}^{\mu\nu}_{2,4} = g^{\mu\nu} \mathcal{O}_{F,m}$.
\end{enumerate}
The renormalisation constants are subject to a constraint following from the non-renormalisation \cite{Freedman:1974gs,Adler:1976zt,Collins:1976yq,Nielsen:1977sy,Collins:1994ee} of the EMT. Since we have allowed for non-$\MSbar$ renormalisation constants, the equality of the renormalised and bare EMTs is not guaranteed anymore. However, we further {\bf assume} that this equality holds:
\begin{empheq}[box=\fbox]{equation} \label{eq:EMTnonRenormalisation}
    \big[ \Theta ]_\mathrm{R} \equiv \big[ \Theta \big]_\mathrm{B} \; .
\end{empheq}
As a consequence, we obtain the constraints:
\begin{equation} \label{eq:EMTnonRenormalisationConstraint} 
\begin{aligned}
    Z_{13} &= 1 - Z_{33} \; , \qquad & Z_{31} &= 1 - Z_{11} \; , \\[.2cm]
    Z_{32} &= \frac{1}{4} - \frac{1}{4} Z_{FF} - Z_{12} \; , \qquad & Z_{34} &= - \frac{1}{4} Z_{Fm} - Z_{14} \; .
\end{aligned}
\end{equation}

It turns out that the renormalisation constants that we have allowed to be defined in a non-$\MSbar$ scheme, $Z_{ij}$, $i=1,3$, $j=2,4$, may be further constrained by considering a different operator basis containing traceless rank-two operators that correspond to an irreducible representation of the Lorentz group. With the following notation for the trace of a rank-two tensor operator:
\begin{equation}
    \Tr(\mathcal{A}) \equiv \mathcal{A}^\rho{}_\rho \; ,
\end{equation}
there is:
\begin{equation} \label{eq:traces}
    \Tr(\mathcal{O}_1) = -\mathcal{O}_F \; , \qquad
    \Tr(\mathcal{O}_2) = D \mathcal{O}_F \; , \qquad
    \Tr(\mathcal{O}_3) = \mathcal{O}_m \; , \qquad
    \Tr(\mathcal{O}_4) = D \mathcal{O}_m \; .
\end{equation}
where the space-time dimension is denoted as $D=4 - 2 \epsilon$ in dimensional regularization.~\\
Let us define:
\begin{equation}
    \tilde{\mathcal{O}}^{\mu\nu}_{1,3} \equiv \mathcal{O}^{\mu\nu}_{1,3} - \frac{g^{\mu\nu}}{D} \Tr(\mathcal{O}_{1,3}) \; . 
\end{equation}
These operators are by construction traceless:
\begin{equation}
    \Tr(\tilde{\mathcal{O}}_{1,3}) = 0 \; .
\end{equation}
We extend the set by rescaling the remaining oprators so that:
\begin{equation}
    \Tr(\tilde{\mathcal{O}}_{2}) = \mathcal{O}_F \; , \qquad
    \Tr(\tilde{\mathcal{O}}_{4}) = \mathcal{O}_m \; .
\end{equation}
The operators $\tilde{\mathcal{O}}_{1,\dots,4}$ are related to $\mathcal{O}_{1,\dots,4}$ by:
\begin{equation}
    \tilde{\mathcal{O}}_i \equiv M_{ij} \mathcal{O}_j \; , \qquad M \equiv \mqty(1 & 1/D & 0 & 0 \\[.2cm] 0 & 1/D & 0 & 0 \\[.2cm] 0 & 0 & 1 & -1/D \\[.2cm] 0 & 0 & 0 & 1/D ) \; , \qquad M^{-1} = \mqty(1 & -1 & 0 & 0 \\[.2cm] 0 & D & 0 & 0 \\[.2cm] 0 & 0 & 1 & 1 \\[.2cm] 0 & 0 & 0 & D ) \; ,
\end{equation}
while their renormalisation takes the form:
\begin{equation}
    \big[ \tilde{\mathcal{O}}_i ]_\mathrm{R} \equiv \tilde{Z}_{ij} \big[ \tilde{\mathcal{O}}_j \big]_\mathrm{B} \; ,
\end{equation}
with the renormalisation constants defined in the $\MSbar$ scheme:

\begin{equation}
\tilde{Z} = \mqty(Z_{11} & 0      & Z_{13} & 0 \\[.2cm]
                  0      & Z_{FF} & 0      & Z_{Fm} \\[.2cm]
                  Z_{31} & 0      & Z_{33} & 0      \\[.2cm]
                  0      & 0      & 0      & 1 ) \; .
\end{equation}
This structure follows from:
\begin{enumerate}
    \item $\tilde{Z}_{ij} = 0$ for $i = 1,3$ and $j = 2,4$ because $\tilde{\mathcal{O}}_{2,4}$ are proportional to scalar operators, while $\tilde{\mathcal{O}}_{1,3}$ are irreducible rank-two tensor operators;
    \item $\tilde{Z}_{ij} = Z_{ij}$ for $i,j = 1,3$ because $\mathcal{O}_{1,3}$ contain the irreducible operators $\tilde{\mathcal{O}}_{1,3}$ with unit coefficient;
    \item $\tilde{Z}_{ij} = Z_{ij}$ for $i,j = 2,4$ because $\tilde{\mathcal{O}}_{2,4} = 1/D \, \mathcal{O}_{2,4}$.
\end{enumerate}
With the help of $\tilde{\mathcal{O}}_{1,\dots,4}$, it is possible to define another set of renormalised operators that are equal to $\mathcal{O}_{1,\dots,4}$ at tree-level:
\begin{equation}
    \big[ \mathcal{O}'_i \big]_\mathrm{R} \equiv M_{ij}^{-1} \big[ \tilde{\mathcal{O}}_j \big]_\mathrm{R} \; .
\end{equation}
The difference between the two bases is:
\begin{equation}
    \big[ \mathcal{O}_i \big]_\mathrm{R} - \big[ \mathcal{O}'_i \big]_\mathrm{R} = \big( \mathbbm{1} - M^{-1} \tilde{Z} M Z^{-1} \big)_{ij} \big[ \mathcal{O}_j \big]_\mathrm{R} \; .
\end{equation}
Explicit calculation confirms the expectation that the difference vanishes for $i=2,4$, while it is proportional to $\big[ \mathcal{O}_{2,4} \big]_\mathrm{R}$ for $i=1,3$:
\begin{equation}
    \big[ \mathcal{O}_{2,4} \big]_\mathrm{R} - \big[ \mathcal{O}'_{2,4} \big]_\mathrm{R} = 0 \; , 
\end{equation}
\begin{equation} \label{eq:xyDefs}
    \big[ \mathcal{O}_{1,3} \big]_\mathrm{R} - \big[ \mathcal{O}'_{1,3} \big]_\mathrm{R} = \frac{1}{D} \Big(  x_{1,3} \big[ \mathcal{O}_2 \big]_\mathrm{R} + y_{1,3} \big[ \mathcal{O}_4 \big]_\mathrm{R} \Big) \; ,
\end{equation}
where:
\begin{equation}  \label{eq:xyz}
\begin{aligned}
    x_1 &= \frac{Z_{FF}-Z_{11}+D Z_{12}}{Z_{FF}} \; , \qquad
    &y_1 &= (1 - x_1) Z_{Fm} + Z_{13} + D Z_{14} \; , \\[.2cm]
    x_3 &= \frac{-Z_{31}+D Z_{32}}{Z_{FF}} \; , \qquad
    &y_3 &= -1 - x_3 Z_{Fm} + Z_{33} + D Z_{34} \; .
\end{aligned}
\end{equation}
Due to Eqs.~\eqref{eq:EMTnonRenormalisationConstraint}, only two of the constants $x_{1,3}, y_{1,3}$ are independent:
\begin{equation}
    x_1 + x_3 = \frac{\epsilon}{2} \big( 1 - Z_{FF}^{-1} \big) = \frac{\beta}{2} \; , \qquad y_1 + y_3 = \frac{\epsilon}{2} Z_{Fm} Z_{FF}^{-1} = -2\gamma_m \; ,
\end{equation}
where we have also used Eqs.~\eqref{eq:Spiridionov} to obtain the right-hand sides of the relations. The constants $x_{1,3}, y_{1,3}$  vanish at tree-level and are regular at $\epsilon = 0$:
\begin{equation}
     x_{1,3}, y_{1,3} = \order{a_s} \; , \qquad x_{1,3}, y_{1,3} \equiv x^{(0)}_{1,3}, y^{(0)}_{1,3} + \order{\epsilon} \; , \qquad x^{(0)}_{1,3}, y^{(0)}_{1,3} \in \mathbb{R} \; .
\end{equation}
Furthermore, they provide a parameterisation of $Z_{ij}$, $i=1,3$, $j=2,4$:
\begin{equation} \label{eq:zyx} 
\begin{aligned}
    Z_{12} &= \frac{1}{D} \big( Z_{11} + (x_1 - 1) Z_{FF} \big) \; , \qquad
    &Z_{14} &= \frac{1}{D} \big( y_1 - Z_{13} + (x_1 - 1) Z_{Fm} \big) \; ,  \\[.2cm]
    Z_{32} &= \frac{1}{D} \big( Z_{31} + x_3 Z_{FF} \big) \; , \qquad
    &Z_{34} &= \frac{1}{D} \big( 1 + y_3 - Z_{33} + x_3 Z_{Fm} \big) \; .
\end{aligned} 
\end{equation}

\subsection{$\MSbar$ scheme constraints} \label{sec:MSbarConstraints}

The case of $Z_{ij}$, $i=1,3$, $j=2,4$ defined in the $\MSbar$ scheme provides further simplifications. Defining:
\begin{equation}
    Z_{ij}(a_s,\epsilon) \equiv \delta_{ij} + \sum_{l=1}^\infty \frac{Z_{ij}^{(l)}(a_s)}{\epsilon^l} \; .
\end{equation}
Eqs.~\eqref{eq:xyz} yield at $\order{\epsilon^0}$:
\begin{equation} \label{eq:xyMSbar}
     x_1^{\MSbar} = -2 Z^{(1)}_{12} \; , \qquad
     y_1^{\MSbar} = -2 Z^{(1)}_{14} \; , \qquad
     x_3^{\MSbar} = -2 Z^{(1)}_{32} \; , \qquad
     y_3^{\MSbar} = -2 Z^{(1)}_{34} \; .
\end{equation}
The same equations \eqref{eq:xyz} taken at $\order{\epsilon^k}$, $k < 0$ provide constraints on $Z_{ij}$, $i=1,3$, $j=2,4$ that determine these constants uniquely. The results may be expressed through the elements of the  anomalous-dimension matrix of $\big[ \mathcal{\tilde{O}}_{1,3} \big]_\mathrm{R}$:
\begin{equation} \label{eq:gamma11and33}
    \dv{\ln \mu^2} \mqty( \big[ \mathcal{\tilde{O}}_1 \big]_\mathrm{R} \\[.2cm] \big[ \mathcal{\tilde{O}}_3 \big]_\mathrm{R} ) = \mqty( \gamma_{11} & -\gamma_{33} \\[.2cm] -\gamma_{11} & \gamma_{33} ) \mqty( \big[ \mathcal{\tilde{O}}_1 \big]_\mathrm{R} \\[.2cm] \big[ \mathcal{\tilde{O}}_3 \big]_\mathrm{R} ) \; .
\end{equation}
Notice that we have exploited Eqs.~\eqref{eq:EMTnonRenormalisationConstraint} to simplify the form of the anomalous-dimension matrix and express the scale dependence of the operators with the help of only two independent anomalous dimensions, $\gamma_{11},\gamma_{33}$. With the expansions:
\begin{equation}
    \beta \equiv -\sum_{l=1}^\infty a_s^l \beta_{l-1} \; , \qquad
    \gamma_m \equiv \sum_{l=1}^\infty a_s^l \gamma_m^{(l)} \; , \qquad
    \gamma_{ij} \equiv \sum_{l=1}^\infty a_s^l \gamma_{ij}^{(l)} \; ,
\end{equation}
there is for instance \cite{Hatta:2018sqd,Tanaka:2018nae}:
\begin{equation} \label{eq:xyMSbarExplicit}
\begin{split}
    x_1^{\MSbar} &= \frac{a_s}{2} \big( - \beta_0 + \gamma_{11}^{(1)} \big) \\[.2cm]
 &\quad + \frac{a_s^2}{8} \big( -4 \beta_1 + \big( \beta_0 - \gamma_{11}^{(1)} - \gamma_{33}^{(1)} \big) \, \gamma_{11}^{(1)} + 2 \gamma_{11}^{(2)} \big) + \order{a_s^3} \\[.2cm]
 &= a_s \, \Big( -\frac{11}{6} C_A \Big) + \order{a_s^2} \; , \\[.4cm]
    y_1^{\MSbar} &= \frac{a_s}{2} \big( - 4 \gamma_m^{(1)} + \gamma_{33}^{(1)} \big) \\[.2cm]
 &\quad + \frac{a_s^2}{8} \big( - 16 \gamma_m^{(2)} + 8 \gamma_{11}^{(1)} \gamma_m^{(1)} - \big( \beta_0 + \gamma_{11}^{(1)} + \gamma_{33}^{(1)} \big) \, \gamma_{33}^{(1)} + 2 \gamma_{33}^{(2)} \big) + \order{a_s^3} \\[.2cm]
 &= a_s \, \Big( \frac{14}{3} C_F \Big) + \order{a_s^2} \; .
 \end{split}
\end{equation}
The quadratic Casimir color constants are defined as usual: $C_A = N_c \,, \, C_F = (N_c^2 - 1)/(2 N_c) \,$ with the number of colors in the fundamental representation $N_c =3$ in QCD.

\section{Quark and gluon EMTs with Renormalisation-Group-invariant traces} \label{sec:RGIqgEMTtraces}

Let us now define renormalised quark and gluon EMTs by decomposing the EMT according to the fields occurring in the operators similarly to Refs.~\cite{Hatta:2018sqd,Tanaka:2018nae}:
\begin{empheq}[box=\fbox]{gather}
    \big[ \Theta ]_\mathrm{R} = \Big[ \mathcal{O}_1 + \frac{1}{4} \mathcal{O}_2 \Big]_\mathrm{R} + \big[ \mathcal{O}_3 \big]_\mathrm{R} \equiv \Big[ \Theta_g \big]_\mathrm{R} + \big[ \Theta_q \big]_\mathrm{R} \; , \\[.2cm]
    \big[ \Theta_g ]_\mathrm{R} = \Big[ \mathcal{O}_1 + \frac{1}{4} \mathcal{O}_2 \Big]_\mathrm{R} \; , \qquad
    \big[ \Theta_q ]_\mathrm{R} = \big[ \mathcal{O}_3 \big]_\mathrm{R} \; , \label{eq:EMTgq}
\end{empheq}
where the renormalised operators on the r.h.s.\ are defined by Eqs.~\eqref{eq:OiRenormalisation}, \eqref{eq:ZijStructure} and \eqref{eq:zyx}. The constants $x_{1,3}, y_{1,3}$ that enter the definitions may in principle have a non-trivial dependence on $\epsilon$. In the $\MSbar$ scheme, they are pure numbers, see Eqs.~\eqref{eq:xyMSbar}. According to Eqs.~\eqref{eq:xyDefs}:
\begin{equation} \label{eq:qgEMTsRelativeToMSbar}
    \big[ \Theta_{g,q} ]_\mathrm{R} = \big[ \Theta_{g,q} ]_\mathrm{R}^{\MSbar} + \frac{1}{D} \Big(  \big( x_{1,3}-x_{1,3}^{\MSbar} \big) \big[ \mathcal{O}_2 \big]_\mathrm{R} + \big( y_{1,3}-y_{1,3}^{\MSbar} \big) \big[ \mathcal{O}_4 \big]_\mathrm{R} \Big) \; ,
\end{equation}
and the differences $x_{1,3}-x_{1,3}^{\MSbar}$ and $y_{1,3}-y_{1,3}^{\MSbar}$ only influence physical observables at $\epsilon = 0$. Hence, as far as applications are concerned, we may assume:
\begin{equation}
    x_{1,3}, y_{1,3} \in \mathbb{R} \; .
\end{equation}
In this Section, we work exclusively with renormalised operators. In consequence, we set $\epsilon = 0$ in all expressions.

Our goal is to obtain a decomposition of the EMT for which both $\Theta_g$ and $\Theta_q$ have RG-invariant traces. This allows to attach a physical meaning to matrix elements of these traces. The RG invariance of the trace of $\Theta$ itself is a consequence of Eq.~\eqref{eq:EMTnonRenormalisation}. Using Eqs.~\eqref{eq:traces}, \eqref{eq:OFOmRenormalisation} and \eqref{eq:Spiridionov}, we recover the well-known result:
\begin{equation} \label{eq:EMTtrace}
    \Tr(\big[ \Theta \big]_\mathrm{R}) = \Tr(\Theta) = \frac{\beta}{2} \big[ \mathcal{O}_F \big]_\mathrm{R} + \big( 1 - 2 \gamma_m \big) \big[ \mathcal{O}_m \big]_\mathrm{R} \; .
\end{equation}
The crucial observation is that the traces of the quark and gluon EMTs will also be expressed through $\big[ \mathcal{O}_F \big]_\mathrm{R}$ and $\big[ \mathcal{O}_m \big]_\mathrm{R}$ which may be combined linearly to yield two RG-invariant operators:
\begin{equation} \label{eq:RGInvariants}
    \dv{\ln \mu^2} \Tr(\Theta) = 0 \; , \qquad
    \dv{\ln \mu^2} \big[ \mathcal{O}_m \big]_\mathrm{R} = 0 \; .
\end{equation}
In terms of these invariant operators:
\begin{equation} \label{eq:qgEMTtracesGeneral}
\begin{split}
    \Tr(\big[ \Theta_g \big]_\mathrm{R}) &= x_1 \big[ \mathcal{O}_F \big]_\mathrm{R} + y_1 \big[ \mathcal{O}_m \big]_\mathrm{R} \\[.2cm]
    &= \frac{2x_1}{\beta} \Tr(\Theta) + \Big( y_1 - \frac{2 x_1}{\beta} \big( 1-2\gamma_m \big) \Big) \big[ \mathcal{O}_m \big]_\mathrm{R} \; , \\[.4cm]
    \Tr(\big[ \Theta_q \big]_\mathrm{R}) &= x_3 \big[ \mathcal{O}_F \big]_\mathrm{R} + \big( 1 + y_3 \big) \big[ \mathcal{O}_m \big]_\mathrm{R} \\[.2cm]
    &= \frac{2x_3}{\beta} \Tr(\Theta) + \Big( 1 + y_3 - \frac{2 x_3}{\beta} \big( 1-2\gamma_m \big) \Big) \big[ \mathcal{O}_m \big]_\mathrm{R} \; .
\end{split}
\end{equation}
Because the scale dependence of $a_s$ is $\order{a_s^2}$, the scale dependence of the coefficients of the invariant operators on the r.h.s.\ is $\order{a_s^2}$ independently of the particular choice of the constants $x_{1,3},y_{1,3}$. Hence, the scale dependence of the traces is always $\order{a_s^2}$. 
Let us now state the conditions that lead to RG-invariant traces:

\vspace{.5cm}
\noindent
\fbox{
\begin{minipage}{.965\textwidth}
    if the constants $x_{1,3}, y_{1,3}$ that define the renormalised EMTs Eqs.~\eqref{eq:EMTgq} through the renormalisation constants Eqs.~\eqref{eq:zyx} satisfy:
    \begin{equation} \label{eq:x0y0deq}
        \dv{\ln \mu^2} \Big( \frac{x_1}{\beta} \Big) = 0 \; , \qquad
        \dv{y_1}{\ln \mu^2} = - \frac{4 x_1}{\beta} \dv{\gamma_m}{\ln \mu^2} = - 4 x_1 \, a_s \dv{\gamma_m}{a_s} \; ,
    \end{equation}
    \begin{equation}
        x_3 = \frac{\beta}{2} - x_1 \; , \qquad y_3= -2\gamma_m - y_1 \; ,
    \end{equation}
    then the traces of the renormalised quark and gluon EMTs are RG invariant:
    \begin{equation}
        \dv{\ln \mu^2} \Tr(\big[ \Theta_{g,q} \big]_\mathrm{R}) = 0 \; .
    \end{equation}
\end{minipage}}
\vspace{.5cm}

\noindent
The constants $x_{1,3}, y_{1,3}$ that yield RG-invariant quark and gluon EMT traces are thus some very general series in $a_s(\mu_0)$ starting from $\order{a_s(\mu_0)}$, with their dependence on the scale $\mu^2$ determined by solving Eqs.~\eqref{eq:x0y0deq}. This introduces an explicit dependence on $\ln \mu^2/\mu_0^2$ given that:
\begin{equation}
    \dv{\ln \mu^2} = \pdv{\ln \mu^2} + \beta a_s \pdv{a_s} \; .
\end{equation}
The form of $x_{1,3}, y_{1,3}$ is substantially constrained, if we further assume that they do not explicitly depend on the scale $\mu$, but are at most functions of $a_s$. Indeed, in this case:
\begin{equation} \label{eq:RGIx}
    x_1 = \frac{\beta}{2} \, k \; , \qquad
    x_3 = \frac{\beta}{2} \big( 1 - k \big) \; , \qquad k \in \mathbb{R} \; .
\end{equation}
The general solution to the second of Eqs.~\eqref{eq:x0y0deq} is $y_1 = - 2 \gamma_m \, k + c$, $c \in \mathbb{R}$. However, since $y_1$ is $\order{a_s}$, $c$ must vanish:
\begin{equation} \label{eq:RGIy}
    y_1 = - 2 \gamma_m \, k \; , \qquad
    y_3 = - 2 \gamma_m \, \big( 1 - k \big) \; .
\end{equation}
This yields the traces:
\begin{equation} \label{eq:qgEMTtraces}
    \begin{split}
    \Tr(\big[ \Theta_g \big]_\mathrm{R}) 
    &= k \, \big( \Tr(\Theta) - \big[ \mathcal{O}_m \big]_\mathrm{R} \big) = k \, \Big( \frac{\beta}{2} \big[ \mathcal{O}_F \big]_\mathrm{R} - 2 \gamma_m \big[ \mathcal{O}_m \big]_\mathrm{R} \Big) \; ,  \\[.4cm]
    \Tr(\big[ \Theta_q \big]_\mathrm{R}) 
    &= \big( 1 - k \big) \big( \Tr(\Theta) - \big[ \mathcal{O}_m \big]_\mathrm{R} \big) + \big[ \mathcal{O}_m \big]_\mathrm{R} \; .
    \end{split}
\end{equation}
Let us finally note that Eqs.~\eqref{eq:qgEMTtraces} cover, in particular, various special cases proposed previously in the literature. For instance, $k = 1$ corresponds to the D2 scheme of Ref.~\cite{Metz:2020vxd}, while $k = 0$ corresponds to the D3 scheme of Ref.~\cite{Lorce:2021xku}.

\subsection{Perturbative correction of non-RG-invariant traces} \label{sec:minimalRGIqgEMTtraces}

Suppose that we are given a decomposition of the EMT Eq.~\eqref{eq:EMTgq} characterised by some $x_{1,3},y_{1,3}$ that are functions of $a_s$ only but do not match at least one of Eqs.~\eqref{eq:RGIx}, \eqref{eq:RGIy}. An example would be the $\MSbar$ constants Eqs.~\eqref{eq:xyMSbarExplicit}. In this case, the traces Eqs.~\eqref{eq:qgEMTtracesGeneral} are not RG invariant:
\begin{equation}
    \dv{\ln \mu^2} \mqty( \Tr(\big[ \Theta_g \big]_\mathrm{R}) \\[.2cm] \Tr(\big[ \Theta_q \big]_\mathrm{R})) = \mqty( \gamma^{\mathrm{tr}}_{gg} & -\gamma^{\mathrm{tr}}_{qq} \\[.2cm] -\gamma^{\mathrm{tr}}_{gg} & \gamma^{\mathrm{tr}}_{qq} ) \mqty( \Tr(\big[ \Theta_g \big]_\mathrm{R}) \\[.2cm] \Tr(\big[ \Theta_q \big]_\mathrm{R})) \equiv \gamma^{\mathrm{tr}} \mqty( \Tr(\big[ \Theta_g \big]_\mathrm{R}) \\[.2cm] \Tr(\big[ \Theta_q \big]_\mathrm{R})) \; , \qquad \gamma^{\mathrm{tr}} \neq 0 \; ,
\end{equation}
where we have used the invariance of the trace of the EMT, Eq.~\eqref{eq:RGInvariants}. 
As noted above, the matrix elements of $\gamma^{\mathrm{tr}}$ are always $\order{a_s^2}$, independently of the particular $x_{1,3},y_{1,3}$.
One can, nevertheless, perform a rotation of the traces to obtain RG-invariant expressions:
\begin{equation}
    \mqty( \Tr(\big[ \Theta_g' \big]_\mathrm{R}) \\[.2cm] \Tr(\big[ \Theta_q' \big]_\mathrm{R})) \equiv F \, \mqty( \Tr(\big[ \Theta_g \big]_\mathrm{R}) \\[.2cm] \Tr(\big[ \Theta_q \big]_\mathrm{R})) \; , \qquad \dv{F}{\ln \mu^2} = - F \, \gamma^{\mathrm{tr}} \; .
\end{equation}
Assuming that $F$ is perturbative, i.e.\ $F(a_s = 0) = \mathbbm{1}$, it can be uniquely determined by solving the above RG equation for any given $x_{1,3},y_{1,3}$. Furthermore, the 
result of the rotation must necessarily match the form of Eqs.~\eqref{eq:qgEMTtraces} with:
\begin{empheq}[box=\fbox]{equation} \label{eq:MinimalCorrection}
    k = \eval{\frac{2 x_1}{\beta}}_{a_s = 0} \; .
\end{empheq}
Indeed, there is for instance:
\begin{equation}
    \Tr(\big[ \Theta_g \big]_\mathrm{R}) - \Tr(\big[ \Theta_g' \big]_\mathrm{R}) = \Big( \frac{2x_1}{\beta} - k \Big) \Tr(\Theta) + \Big( y_1 - \frac{2 x_1}{\beta} \big( 1 - 2\gamma_m \big) + k \Big) \big[ \mathcal{O}_m \big]_\mathrm{R} \; .
\end{equation}
However, $F = \mathbbm{1} + \order{a_s}$ implies:
\begin{equation}
    \frac{2x_1}{\beta} - k = \order{a_s} \; , \qquad y_1 - \frac{2 x_1}{\beta} \big( 1 - 2\gamma_m \big) + k = \order{a_s} \; ,
\end{equation}
which yields Eq.~\eqref{eq:MinimalCorrection}.

The above procedure applied to the D1 scheme of Ref.~\cite{Rodini:2020pis} yields the  D2 scheme of Ref.~\cite{Metz:2020vxd}. In the present publication we make the following proposal:

\vspace{.5cm}
{\centering
\fbox{
\begin{minipage}{0.965\textwidth} Let the quark and gluon EMTs, $[\Theta_{q}]_R^\mathrm{RGItr}$ and $[\Theta_{g}]_R^\mathrm{RGItr}$, with RG-invariant traces be defined by perturbatively correcting the respective tensors defined in the $\MSbar$ scheme. This corresponds to setting in Eqs.~\eqref{eq:qgEMTtraces}:
\begin{equation} \label{eq:PreferredKChoice}
    k = \frac{\beta_0 \big|_{n_f = 0}}{\beta_0} = \Big( 1 - \frac{2}{33} n_f \Big)^{-1} \; , \qquad \beta_0 = \frac{11}{3} C_A - \frac{4}{3} T_F n_f \; ,
\end{equation}
\end{minipage}
}}
\vspace{.5cm}

\noindent
where we have used Eq.~\eqref{eq:xyMSbarExplicit}, and $n_f$ in $\beta_0$ is the number of massless quarks and the color-trace normalization factor is denoted by $T_F$. 
$[\Theta_{g,q}]_R^\mathrm{RGItr}$ are defined by Eqs.~\eqref{eq:EMTgq} with the renormalised operators defined in Eqs.~\eqref{eq:OiRenormalisation}, \eqref{eq:ZijStructure} with non-$\MSbar$-scheme renormalisation constants Eqs.~\eqref{eq:zyx} with $x_{1,3},y_{1,3}$ from Eqs.~\eqref{eq:RGIx} and \eqref{eq:RGIy}.

\section{Example applications at four-loop order} \label{sec:FourLoopResults}

The expectation value of the trace of the EMT in a single-particle state is proportional to the square of the mass, $m$, of the particle:
\begin{equation}
\label{eq:ParticleMass}
    \ev{\Tr(\Theta)}{mj;\bm{p}\lambda} = 2 m^2 \quad \text{for} \quad \ip{mj;\bm{p}\lambda}{mj;\bm{p}'\lambda'} = 2E_{\bm{p}} \big(2\pi\big)^3 \delta^{(3)}(\bm{p} - \bm{p}') \delta_{\lambda\lambda'} \; ,
\end{equation}
where $j$ and $\lambda,\lambda'$ are the spin and polarisations of the particle, $\bm{p},\bm{p'}$ are its three-momenta and $E_{\bm{p}}$ is its energy. The decomposition Eq.~\eqref{eq:EMTgq} of the EMT into quark and gluon EMTs may be used to define the respective quark and gluon contributions to the mass. Unfortunately, this decomposition is not unique. If the occurring operators $\mathcal{O}_{1,\dots,3}$ are defined in the $\MSbar$ scheme, then the quark and gluon contributions to the mass depend on the unphysical renormalisation scale $\mu$. With the renormalisation scheme specified in Section~\ref{sec:RGIqgEMTtraces}, on the other hand, these contributions do not depend on $\mu$ anymore, but there still remains a dependence on the parameter $k$, see Eqs.~\eqref{eq:qgEMTtraces}. This remaining ambiguity is finally removed with our proposal, Eq.~\eqref{eq:PreferredKChoice}. In this Section, we compare the quark and gluon contributions to the mass of the proton and pion defined in the $\MSbar$ scheme to those obtained with our proposal.

We begin with the $\MSbar$-renormalised quark and gluon contributions to the proton mass originally presented in Ref.~\cite{Tanaka:2018nae} at three-loop order. Since the anomalous dimensions, $\gamma_{11}$ and $\gamma_{33}$, of the $\mathcal{O}_{1,3}$ operators, Eq.~\eqref{eq:gamma11and33}, have recently been determined to four-loop order \cite{Moch:2017uml,Moch:2021qrk}, and the QCD $\beta$-function as well as the quark-mass anomalous dimension are available at the same order from Refs.~\cite{vanRitbergen:1997va,Czakon:2004bu} and \cite{Chetyrkin:1997dh,Vermaseren:1997fq}, it is now possible to extend the analysis of Ref.~\cite{Tanaka:2018nae} to one order higher\footnote{The state-of-the-art five-loop results for $\beta$ \cite{Baikov:2016tgj,Herzog:2017ohr,Luthe:2017ttg} and $\gamma_m$ \cite{Baikov:2014qja,Luthe:2016xec,Baikov:2017ujl} cannot be used, since the anomalous dimensions of the $\mathcal{O}_{1,3}$ operators are not known at this order yet.}. To this end, we use four-loop accurate generalisations of Eqs.~\eqref{eq:xyMSbarExplicit} substituted in Eqs.~\eqref{eq:qgEMTtracesGeneral}. Due to the length of the resulting expressions, we refrain from reproducing them explicitly in the text, but rather include them in an ancillary file with this publication.

Let $\ket{\mathrm{P}}$ be a proton state with some (irrelevant) polarisation and three-momentum, and consider QCD with $n_f = 3$ light quarks. The expectation values of the traces of the quark and gluon EMTs read:
\begin{equation}
\begin{split}
\label{eq:ProtonMatrixElement}
\ev{\Tr([\Theta_{g}]_R^{\MSbar})}_{\mathrm{P}}
&=
\langle [O_F]_R\rangle_{\mathrm{P}} \, 
\left(
-0.437676 \, \alpha_s 
-0.261512 \, \alpha_s^2
-0.183827 \, \alpha_s^3
-0.256096 \, \alpha_s^4
\right) \\[.2cm] 
&+
\langle [O_m]_R\rangle_{\mathrm{P}} \,
\left(
0.495149 \, \alpha_s 
+0.776587 \, \alpha_s^2
+0.865492 \, \alpha_s^3
+0.974674 \, \alpha_s^4
\right) \,, \\[.4cm]  
\ev{\Tr([\Theta_{q}]_R^{\MSbar})}_{\mathrm{P}}
&=
\langle [O_F]_R\rangle_{\mathrm{P}} \, 
\left(0.079578 \, \alpha_s 
+0.058870 \, \alpha_s^2 
+0.021604 \, \alpha_s^3 
+0.013675 \, \alpha_s^4
\right) \\[.2cm]
&+
\langle [O_m]_R\rangle_{\mathrm{P}} \,
\left(1 \,+\, 
0.141471 \, \alpha_s 
-0.008235 \, \alpha_s^2
-0.064351 \, \alpha_s^3
-0.065869 \, \alpha_s^4
\right) \,, \\[.2cm]
\end{split}
\end{equation}
where $\langle [O_F]_R\rangle_{\mathrm{P}}$, $\langle [O_m]_R\rangle_{\mathrm{P}}$ denote, respectively, the non-perturbative expectation values of the $\MSbar$-renormalised operators $O_F$ and $O_m$. Both $\alpha_s$ and $\langle [O_F]_R\rangle_{\mathrm{P}}$ depend on the renormalisation scale $\mu$, while $\langle [O_m]_R\rangle_{\mathrm{P}}$ is independent of $\mu$ since $[O_m]_R$ is RG invariant. Eqs.~\eqref{eq:ProtonMatrixElement} agree with Ref.~\cite{Tanaka:2018nae} up to $\mathcal{O}(\alpha_s^3)$, while the $\mathcal{O}(\alpha_s^4)$ terms are new. These results may also be rewritten in terms of the expectation values of RG-invariant operators using $\ev{\Tr(\Theta)}_{\mathrm{P}} = 2m_{\mathrm{P}}^2$ in Eqs.~\eqref{eq:qgEMTtracesGeneral}:
\begin{equation}
\begin{split}
\label{eq:ProtonMatrixElement2}
\frac{\ev{\Tr([\Theta_{g}]_R^{\MSbar})}_{\mathrm{P}}}{2m_\mathrm{P}^2}
&= 1.22222
+ 0.0386426 \, \alpha_s 
- 0.0622081 \, \alpha_s^2
- 0.0945524 \, \alpha_s^3 \\
&+
\frac{\langle [O_m]_R\rangle_{\mathrm{P}}}{2m_\mathrm{P}^2} \,
\left(-1.22222
- 0.321585 \, \alpha_s 
- 0.124903 \, \alpha_s^2
- 0.00921612 \, \alpha_s^3
\right) \,,
\\[.4cm]  
\frac{\ev{\Tr([\Theta_{q}]_R^{\MSbar})}_{\mathrm{P}}}{2m_\mathrm{P}^2}
&= -0.222222
- 0.0386426 \, \alpha_s
+ 0.0622081 \, \alpha_s^2
+ 0.0945524 \, \alpha_s^3 \\
&+
\frac{\langle [O_m]_R\rangle_{\mathrm{P}}}{2m_\mathrm{P}^2} \,
\left(1.22222
+ 0.321585 \, \alpha_s 
+ 0.124903 \, \alpha_s^2
+ 0.00921612 \, \alpha_s^3
\right) \,.
\end{split}
\end{equation}
The scale dependence of the expectation values is now entirely due to the strong coupling constant. In order to illustrate the size of this dependence, we need a numerical estimate for $\langle[O_m]_R\rangle_{\mathrm{P}} / 2m_\mathrm{P}^2$. One possibility is to set it to null by taking the chiral limit as has been done in Ref.~\cite{Tanaka:2018nae}. On the other hand, we may also obtain it from lattice simulations, for example from Tab.\ 1 of Ref.~\cite{Liu:2021gco} (see also Refs.~\cite{Yang:2015uis,XQCD:2013odc}) where it is denoted $f^N_{q\,\mathrm{total}}$ at $n_f = 2+1$:
\begin{equation}
    \frac{\langle [O_m]_R\rangle_{\mathrm{P}}}{2m_\mathrm{P}^2} \approx 0.092 \; .
\end{equation}
As for the strong coupling constant, we make use of the program \textsc{RunDec}~\cite{Chetyrkin:2000yt} with four-loop running and default parameter settings which yields\footnote{The values in eq.(\ref{eq:asv}) were determined using \textsc{RunDec}~\cite{Chetyrkin:2000yt} via solving the RGE equation for $\alpha_s$ at 4-loop accuracy numerically with initial value $\alpha_s = 0.118$ at the Z-boson pole mass.}:
\begin{equation} \label{eq:asv}
    \alpha_s(\mu) \approx
    \begin{cases} 
    0.48 & \text{for} \quad \mu = 1\, \text{GeV} \sim m_{\mathrm{P}} \; , \\[.2cm]
    0.30 & \text{for} \quad \mu = 2\, \text{GeV} \; , \\[.2cm]
    0.80 & \text{for} \quad \mu = 1/\sqrt{2} \, \text{GeV} \; . \\[.2cm]
    \end{cases}
\end{equation}
As can be seen from these numbers, a typically-used scale choice for the lower scale amounting to half of the central scale would certainly be non-perturbative. This is the reason for our unusual choice of $1/\sqrt{2} \, \text{GeV}$. Substitution in Eqs.~\eqref{eq:ProtonMatrixElement2} results in:
\begin{equation} \label{eq:ProtonMatrixElement3}
\begin{split}
\frac{\ev{\Tr([\Theta_{g}]_R^{\MSbar})}_{\mathrm{P}}}{2m_\mathrm{P}^2} &= \begin{cases} 
    1.087 & \text{for} \quad \mu = 1\, \text{GeV} \sim m_{\mathrm{P}} \; , \\[.2cm]
    1.103 & \text{for} \quad \mu = 2\, \text{GeV} \; , \\[.2cm]
    1.021 & \text{for} \quad \mu = 1/\sqrt{2} \, \text{GeV} \; ,
    \end{cases} \\[.4cm]
\frac{\ev{\Tr([\Theta_{q}]_R^{\MSbar})}_{\mathrm{P}}}{2m_\mathrm{P}^2} = 
1-\frac{\ev{\Tr([\Theta_{g}]_R^{\MSbar})}_{\mathrm{P}}}{2m_\mathrm{P}^2} &=
\begin{cases} 
    -0.087 & \text{for} \quad \mu = 1\, \text{GeV} \sim m_{\mathrm{P}} \; , \\[.2cm]
    -0.103 & \text{for} \quad \mu = 2\, \text{GeV} \; , \\[.2cm]
    -0.021 & \text{for} \quad \mu = 1/\sqrt{2} \, \text{GeV} \; . 
    \end{cases}
\end{split}
\end{equation}
The gluon content of the proton is rather stable with respect to the scale choice according to the above results. Indeed, the deviations from the value for the central scale reach only $+ 1.5 \%$ and $- 5.9 \%$ for the upper and lower scales respectively. The quark content, on the other hand, is rather uncertain with deviations ranging from $+ 19 \%$ to $- 74 \%$.

Let us now compare the above results to those obtained from Eqs.~\eqref{eq:qgEMTtraces} and \eqref{eq:MinimalCorrection}. Taking into account the construction presented in Section~\ref{sec:minimalRGIqgEMTtraces}, it should be clear that the desired results can be obtained by simply setting $\alpha_s = 0$ in Eqs.~\eqref{eq:ProtonMatrixElement2}:
\begin{equation} \label{eq:ProtonMatrixElement4}
\begin{split}
\frac{\ev{\Tr([\Theta_{g}]_R^\mathrm{RGItr})}_{\mathrm{P}}}{2m_\mathrm{P}^2}
&= \quad 1.22222 - 1.22222 \, \frac{\langle [O_m]_R\rangle_{\mathrm{P}}}{2m_\mathrm{P}^2}  = \quad 1.11 \,,
\\[.4cm]  
\frac{\ev{\Tr([\Theta_{q}]_R^\mathrm{RGItr})}_{\mathrm{P}}}{2m_\mathrm{P}^2}
&= -0.222222 + 1.22222 \, \frac{\langle [O_m]_R\rangle_{\mathrm{P}}}{2m_\mathrm{P}^2} = -0.11 \,.
\end{split}
\end{equation}
These numbers are, expectedly, very close to those listed in Eqs.~\eqref{eq:ProtonMatrixElement2} for the central scale choice. Their main advantage is that they are scale independent and have the same value even in the non-perturbative regime of the strong coupling constant.

Finally, let us note that if the quark and gluon EMTs were not defined through renormalised operators in Eqs.~\eqref{eq:EMTgq}, their traces would nevertheless be finite and, of course, RG invariant. For reference, we also list the corresponding values:
\begin{equation} \label{eq:ProtonMatrixElement5}
\begin{aligned}
\frac{\ev{\Tr([\Theta_{g}]_B)}_{\mathrm{P}}}{2m_\mathrm{P}^2}
&= 1 - \frac{\langle [O_m]_R\rangle_{\mathrm{P}}}{2m_\mathrm{P}^2} = 0.908 \,,
\\[.4cm]  
\frac{\ev{\Tr([\Theta_{q}]_B)}_{\mathrm{P}}}{2m_\mathrm{P}^2}
&= \qquad \frac{\langle [O_m]_R\rangle_{\mathrm{P}}}{2m_\mathrm{P}^2} = 0.092 \,.
\end{aligned}
\end{equation}
These numbers correspond to the classical EMT trace assigned entirely to the quark EMT, as in Ji's decomposition scheme~\cite{Ji:1994av}, see also Ref.~\cite{Liu:2021gco}. Furthermore, they are identical to those obtained with $k = 1$ in Eq.~\eqref{eq:qgEMTtraces}, i.e.\ with the D2 scheme of Ref.\cite{Metz:2020vxd}. As far as the gluon contribution to the proton mass is concerned, the difference between \eqref{eq:ProtonMatrixElement4} and \eqref{eq:ProtonMatrixElement5} is about 20\%. However, there is a qualitative difference as far as the quark contribution is concerned, because of the sign difference between \eqref{eq:ProtonMatrixElement4} and \eqref{eq:ProtonMatrixElement5}. \eqref{eq:ProtonMatrixElement4} implies that quarks reduce the mass of the proton, while \eqref{eq:ProtonMatrixElement5} implies that quarks increase the mass of the proton.

Let us now turn to a pion state $| \pi \rangle$ with mass $m_{\pi}$. We base our discussion on Ref.~\cite{Tanaka:2018nae}, where chiral perturbation theory has been used to derive the following approximation:
\begin{equation}
\langle [O_m]_R\rangle_{\pi} = m_{\pi}^2\,.
\end{equation}
In this case Eq.~\eqref{eq:qgEMTtracesGeneral} turns into:
\begin{equation} \label{eq:PionMatrixElement}
\begin{split}
\frac{\ev{\Tr([\Theta_{g}]_R)^{\MSbar}}_{\pi}}{2m_\pi^2} &= \big( 1+2\gamma_m \big) \frac{x_1^{\MSbar}}{\beta} + \frac{y_1^{\MSbar}}{2} \\[.2cm]
&= 0.611111 
- 0.122150 \, \alpha_s
- 0.124659 \, \alpha_s^2
- 0.099160 \, \alpha_s^3\,, 
\\[.4cm]
\frac{\ev{\Tr([\Theta_{q}]_R^{\MSbar})}_{\pi}}{2m_\pi^2} &= 1 - \frac{\ev{\Tr([\Theta_{g}]_R^{\MSbar})}_{\pi}}{2m_\pi^2} \\[.2cm]
&= 0.388889
+ 0.122150 \, \alpha_s
+ 0.124659 \, \alpha_s^2
+ 0.099160 \, \alpha_s^3 \,.
\end{split}
\end{equation}
Eqs.~\eqref{eq:PionMatrixElement} agree with the results presented in Ref.~\cite{Tanaka:2018nae} only up to $\mathcal{O}(\alpha_s^2)$. The discrepancy at $\mathcal{O}(\alpha_s^3)$ is due to the presence of $1/\beta$ in the coefficient of $x_1$. Indeed, $x_1$ is therefore needed to $\mathcal{O}(\alpha_s^4)$. At the time of writing of Ref.~\cite{Tanaka:2018nae} this coefficient could not be determined because of the missing terms in the expansions of the anomalous dimensions of the operators $\mathcal{O}_{1,3}$.

We now finally note that at the scale of the pion mass, the strong coupling constant is non-perturbative. In consequence, Eqs.~\eqref{eq:PionMatrixElement} are not particularly useful. On the other hand, in the case of RG-invariant EMT traces, there is:
\begin{equation} \label{eq:PionMatrixElement2}
\begin{split}
\frac{\ev{\Tr([\Theta_{g}]_R^{\mathrm{RGItr}})}_{\pi}}{2m_\pi^2} &= \eval{\frac{x_1^{\MSbar}}{\beta}}_{\alpha_s = 0} = \frac{k}{2} = \frac{\beta_0 \big|_{n_f = 0}}{2\beta_0} = 0.611111 \,, 
\\[.4cm]
\frac{\ev{\Tr([\Theta_{q}]_R^{\mathrm{RGItr}})}_{\pi}}{2m_\pi^2} &= 1 - \frac{\ev{\Tr([\Theta_{g}]_R^{\mathrm{RGItr}})}_{\pi}}{2m_\pi^2} = 0.388889 \,,
\end{split}
\end{equation}
which provides a decomposition of the mass of the pion into quark and gluon components even in the non-perturbative regime. Of course, the exact physical meaning of these numbers without guidance from perturbation theory may still be disputed due to all the ambiguities that we have discussed.

\section{Conclusions} \label{sec:conclusions}

In this publication, we have provided a general construction of quark and gluon EMTs with RG-invariant traces. The only two constraints that we have assumed are: 1) $\MSbar$-renormalisation of the occurring irreducible rank-two tensor operators, 2) equality of the renormalised and the bare total EMT. An alternative construction to the one presented here consists of renormalising all physical operators occurring in the definition of the quark and gluon EMTs, i.e.\ not only the irreducible rank-two tensor operators but also scalar operators, in the $\MSbar$ scheme, followed by the addition of finite contributions proportional to the two renormalised scalar operators. The two approaches are equivalent by Eq.~\eqref{eq:qgEMTsRelativeToMSbar}.

A possible generalisation of our study would consist in allowing for finite contributions to the quark and gluon EMTs proportional to the two renormalised irreducible rank-two tensor operators. This would not change the analysis of the traces of the EMTs, since irreducible tensor operators are traceless. On the other hand, the additional freedom would allow constructing quark and gluon EMTs that are RG invariant themselves rather than only after taking the trace. While such a construction is certainly interesting, it is irrelevant to the physical motivation behind the study of separate quark and gluon EMTs, namely the understanding of the origin of mass of strongly interacting particles. 

Our analysis has shown that the finite coefficients entering the renormalisation of the scalar operators are strongly constrained. Expressing the traces of the EMTs in terms of RG-invariant operators, we have obtained RG equations that these coefficients must satisfy. These equations still allow for rather general solutions. Hence, we have discussed the case where they do not explicitly depend on the renormalisation scale. This case covers the majority of definitions of quark and gluon EMTs presented in the literature and leaves the freedom to choose a single numeric parameter. We have proposed a value for this parameter that corresponds to a minimal modification of quark and gluon EMTs defined in the $\MSbar$ scheme. 
The corresponding operator-renormalisation scheme removes the perturbative correction terms in Eq.~\eqref{eq:ProtonMatrixElement2}, which are consequences of the $\overline{\mathrm{MS}}$ renormalisation of the quark and gluon trace operators. It nevertheless coincides with $\overline{\mathrm{MS}}$ renormalisation at leading order. This feature is desirable in applications to mesons whose mass scales are in the non-perturbative regime of the strong coupling constant. 
As one possible example where the physical analysis may benefit from our operator-renormalisation scheme, one may consider the so-called nucleon's twist-four gravitational form factors in the forward limit, see e.g.\ Refs.~\cite{Ji:1996ek,Lorce:2017xzd,Polyakov:2018zvc,Tanaka:2018wea,Hatta:2018sqd,Tanaka:2018nae,Liu:2021gco}. It is straightforward to prove that the anomalous dimensions in the homogeneous part of the RG equation for these renormalised objects defined in our scheme will be simply those in Eq.~\eqref{eq:gamma11and33}. The latter are known to high orders in perturbation theory from the literature. We hope to return to this point in the future.

Finally, we have updated previous theoretical results for the $\MSbar$-renormalised quark and gluon contributions to the proton and pion mass induced by the respective EMTs to achieve four-loop accuracy.

\section*{Acknowledgements}

We thank G. Bali for useful discussions.
The work of T.A. received funding from the European Research Council (ERC) under the European Union's Horizon 2020 research and innovation programme \textit{High precision multi-jet dynamics at the LHC} (ERC Condsolidator grant agreement No 772009). The work of L.C. was supported by the Natural Science Foundation of China under contract No.12205171, No.12235008. The work of M.C. was supported by the Deutsche Forschungsgemeinschaft (DFG) under grant 396021762 - TRR 257: Particle Physics Phenomenology after the Higgs Discovery.

\bibliography{emt} 
\bibliographystyle{JHEP}
\end{document}